\documentclass[12pt]{article}
\usepackage{amsfonts,latexsym}
\begin{document}
\newcommand{\e}{\epsilon} \newcommand{\ot}{\otimes}
\newcommand{\be}{\begin{equation}} \newcommand{\ee}{\end{equation}}
\newcommand{\ba}{\begin{eqnarray}} \newcommand{\ea}{\end{eqnarray}}
\newcommand{\tmod}{{\cal T}}\newcommand{\amod}{{\cal A}}
\newcommand{\bemod}{{\cal B}}\newcommand{\cmod}{{\cal C}}
\newcommand{\dmod}{{\cal D}}\newcommand{\hmod}{{\cal H}}
\newcommand{\sss}{\scriptstyle}\newcommand{\tr}{{\rm tr}}
\newcommand{\einsop}{{\bf 1}}
\def\oR{R^*} \def\upa{\uparrow}
\def\R{\overline{R}} \def\doa{\downarrow}
\def\nn{\nonumber} \def\dag{\dagger}
\def\be{\begin{equation}}
\def\ee{\end{equation}}
\def\bea{\begin{eqnarray}}
\def\eea{\end{eqnarray}}
\def\ve{\epsilon}
\def\s{\sigma}
\def\th{\theta}
\def\a{\alpha}
\def\b{\beta}
\def\g{\gamma}
\def\h{\overline{h}}
\def\d{\delta}
\def\e{\eta}
\def\p{\Phi}
\def\M{\cal M}
\def\N{\cal N}
\def\u{\overline}
\def\t{\tau}
\def\l{\left}
\def\r{\right}
\newcommand{\reff}[1]{eq.~(\ref{#1})}
%\draft
 \centerline{\large\bf{Solution of a two leg spin }}
\centerline{\large\bf{ladder system  }}
~~~\\
\begin{center}
{\large Jon Links}
\vspace{0.5cm}~~\\
{\em Centre for Mathematical Physics, \\ Department of Mathematics, \\
The University of Queensland,  4072 \\  Australia \\
e-mail jrl@maths.uq.edu.au}
~~\\
~~\\
\vspace{1cm}
{\large Angela Foerster}
\vspace{0.5cm}~~\\
{\em Instituto de F\'{\i}sica da UFRGS, \\
Avenida Bento Gon\c{c}alves 9500, \\
Porto Alegre, \\
RS- Brasil \\
email angela@if.ufrgs.br}
\end{center}
\vspace{1cm}
\begin{abstract}
A new model for a spin 1/2 ladder system with two legs is introduced. It
is demonstrated that this model is solvable via the Bethe ansatz method
for arbitrary values of the rung coupling $J$.
This is achieved by a suitable mapping from the Hubbard model with
appropriate twisted boundary conditions.
% The introduction of the
% twisted boundary conditions is necessary to accommodate for
% the non-locality of the mapping.
We determine that a phase transition
between gapped and gapless spin excitations occurs at the critical value
$J_c=1/2$ of the rung coupling.
\end{abstract}
\vspace{1cm}
\begin{flushleft}
\end{flushleft}

\vfil\eject
%\centerline{{\bf 1. Introduction}}
~~\\

Research in spin ladder systems continues to attract considerable 
attention. This is primarily motivated by the desire to understand 
the phenomenon of high temperature superconductivity observed in doped
antiferromagnetic materials. In studying ladder materials, it is 
anticipated that knowledge of two-dimensional systems can be 
gained by way of extrapolation from the one-dimensional scenario, 
where there exists a greater understanding of the physics from the 
theoretical perspective. Moreover, it is possible to experimentally
study ladder materials and numerical simulations are easier to treat 
than for the two-dimensional case which facilitates a greater interaction 
between theory and phenomenology. For a review of these aspects we refer
to \cite{dr}.

More recently, there has been a push to apply the mathematically rich
techniques of the Quantum Inverse Scattering Method (QISM) \cite{fst}
and associated 
Bethe ansatz procedures, which have successfully been used in the study 
of one-dimensional quantum systems \cite{kbi}, to provide some insight
into the behaviour of the ladder systems. In order to extend the standard
one-dimensional approach of the QISM to the case of ladders, a number 
of approaches have thus far been proposed.

Recall that central to the QISM is the Yang-Baxter equation, solutions of
which, known as $R$-matrices, provide a systematic means to produce 
a quantum system with an infinite number of conserved charges which
guarantees integrability. In the algebraic formulation, there is an 
associated algebra known as a Yangian and in this approach the Bethe ansatz
is described in terms of the representation theory of this algebra. The
work of Frahm and R\"odenbeck \cite{fr} developed a construction for 
ladder systems where the extension from the one-dimensional system 
to the ladder was obtained by an algebra homomorphism of the 
Yangian algebra. In this manner, the symmetry algebra of the ladder
system remains the same as the original one-dimensional model. Closely
related to this approach is that adopted by Muramoto and Takahashi
\cite{mt} who employed the higher order conservation laws obtained
through the QISM to define a two-leg system which generalizes the
long celebrated Majumdar-Ghosh zigzag ladder model \cite{mg}.
 Such 
zigzag systems can also be obtained via an {\it alternating spin}  
formulation such as that considered by Abad and Rios 
\cite{ar} for the $t-J$ model.

Alternatively, the approach can be considered where the symmetry algebra
is extended to describe the ladder model. This notion was promoted by Wang
\cite{w} who successfully constructed a two-leg ladder system based on the
symmetry algebra $su(4)$ as opposed to the $su(2)$ symmetry of the 
one-dimensional Heisenberg chain. A great advantage in employing this 
method is that it allows for the introduction of rung interactions 
of arbitrary coupling strength which in turn allows for the identification
of the phase diagram in terms of this parameter. The rung interaction is
introduced by way of a chemical potential like term. In Wang's analysis, two
critical values of the rung interaction were found indicating three phases,
one of which exhibited gapped spin excitations. Subsequently, this 
method was extended by Batchelor and Maslen \cite{bm1} who determined 
a class 
of $n$-leg systems each with a corresponding $gl(2^n)$ symmetry and 
has also been generalized to some other cases \cite{f,fk,osp}.

Yet another direction was taken by Albeverio et. al. \cite{afw} who searched
for an $R$-matrix solution which was not previously associated with 
a known one-dimensional system. They were able to determine such a solution
where the rung interactions appear in the Hamiltonian directly from its 
derivation from the $R$-matrix. They could then generalize this system
to one with two free interaction parameters similarly to Wang's solution
discussed above. A subsequent analysis of this model has been undertaken
in \cite{bm2}. 

Our aim in this work is to demonstrate that it is possible to
obtain a solvable ladder system with arbitrarily coupled rung interactions 
directly from an $R$-matrix solution. To achieve this end, we begin with 
the coupled spin formulation of the Hubbard model as introduced by Shastry
\cite{s}, on a closed lattice with twisted boundary conditions. The algebraic
Bethe ansatz solution of this model has been studied by Martins and Ramos 
\cite{mr}. By means of carefully chosen transformations, we map this model
to a spin ladder system with periodic boundary conditions. Remarkably,
the resulting model assumes a simple form with three basic forms of 
interaction. The energy expression in terms of a Bethe ansatz solution
is also obtained. For this model, the rung interactions are not of the 
chemical potential type referred to above. Rather, the 
rung interaction parameter appears explicitly in the Bethe ansatz equations
and so it is reasonable to expect that the behaviour of this model to differ
from the class of ladder models with chemical potential type rung 
interaction. We find   the 
critical value $J_c=1/2$ for the rung interaction parameter
indicating the transition between gapped and gapless phases.

Here we show solvability of  the following two-leg ladder Hamiltonian 
with an even number of rungs and periodic boundary conditions. Explicitly,
the global Hamiltonian is of the form
\be H=\sum_{i=1}^{L-1} h_{i(i+1)} + h_{L1} \ee 
where the local Hamiltonians read  
\be h_{ij}=\l(\s^+_i\s^-_j+\s^-_i\s^+_j\r)\l(\t^z_i\t^z_j\r)^{i+1}
+\l(\t^+_i\t^-_j+\t^-_i\t^+_j\r) 
\l(\s^z_i\s^z_j\r)^{i} +J/2 \l({\vec \s}_i.{\vec \t}_i
+{\vec \s}_j.{\vec\t}_j\r).\label{ham} \ee 
Above, the coupling $J$ can take arbitrary values.

The ladder system is depicted graphically below.
Across the rungs there is the usual Heisenberg $XXX$ interaction while
along the legs the interactions alternate between pure and
correlated $XX$ exchanges. Clearly the correlated exchange is a four body
interaction depending on the spins of the opposing leg. 
In the $J=0$ limit the correlated exchanges have no real physical significance
since for this case the model may be mapped back to
two decoupled $XX$ (or free fermion) chains 
with twisted boundary conditions. This is in some contrast to the case of 
\cite{w} where in the absence of rung interactions the model maintains
non-trival interactions between the legs. In the
thermodynamic limit, the boundary conditions become irrelevant and we
conclude that this region is gapless. On the other hand 
for large $J$ the situation is the
same as the two-leg Heisenberg ladder. In this limit the ground state
consists of a product of rung singlets and the excitations are gapped
\cite{dr}. Hence we expect there to exist a finite critical value of $J$
defining the phase transition.
  \vfil\eject
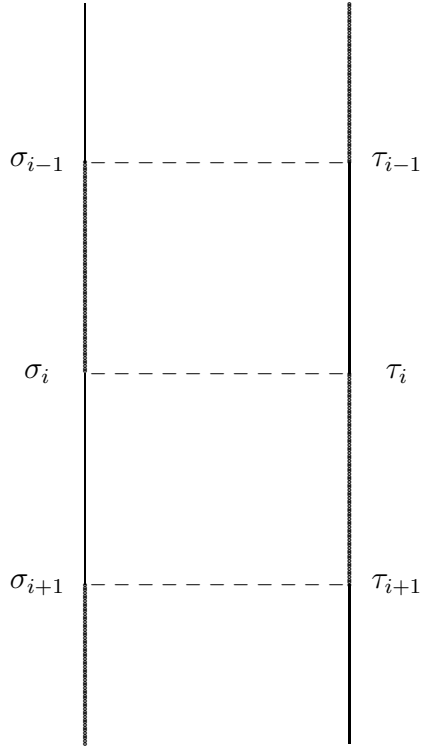
\vspace{1cm}

\begin{picture}(100.,150.)
\put(82.,80.){\makebox(0,0)[cc]{$ \sigma_{i-1}$}}
\put(218.,80.){\makebox(0,0)[cc]{$ \tau_{i-1}$}}
\put(82.,0.){\makebox(0,0)[cc]{$ \sigma_i$}}
\put(218.,0.){\makebox(0,0)[cc]{$ \tau_i$}}
\put(82.,-80.){\makebox(0,0)[cc]{$ \sigma_{i+1}$}}
\put(218.,-80.){\makebox(0,0)[cc]{$ \tau_{i+1}$}}
\put(105.,-80.){\makebox(0,0)[cc]{$ \sss - $}}
\put(114.,-80.){\makebox(0,0)[cc]{$ \sss - $}}
\put(123.,-80.){\makebox(0,0)[cc]{$ \sss - $}}
\put(132.,-80.){\makebox(0,0)[cc]{$ \sss - $}}
\put(141.,-80.){\makebox(0,0)[cc]{$ \sss - $}}
\put(150.,-80.){\makebox(0,0)[cc]{$ \sss - $}}
\put(159.,-80.){\makebox(0,0)[cc]{$ \sss - $}}
\put(168.,-80.){\makebox(0,0)[cc]{$ \sss - $}}
\put(177.,-80.){\makebox(0,0)[cc]{$ \sss - $}}
\put(186.,-80.){\makebox(0,0)[cc]{$ \sss - $}}
\put(195.,-80.){\makebox(0,0)[cc]{$ \sss - $}}

\put(105.,0.){\makebox(0,0)[cc]{$ \sss - $}}
\put(114.,0.){\makebox(0,0)[cc]{$ \sss - $}}
\put(123.,0.){\makebox(0,0)[cc]{$ \sss - $}}
\put(132.,0.){\makebox(0,0)[cc]{$ \sss - $}}
\put(141.,0.){\makebox(0,0)[cc]{$ \sss - $}}
\put(150.,0.){\makebox(0,0)[cc]{$ \sss - $}}
\put(159.,0.){\makebox(0,0)[cc]{$ \sss - $}}
\put(168.,0.){\makebox(0,0)[cc]{$ \sss - $}}
\put(177.,0.){\makebox(0,0)[cc]{$ \sss - $}}
\put(186.,0.){\makebox(0,0)[cc]{$ \sss - $}}
\put(195.,0.){\makebox(0,0)[cc]{$ \sss - $}}

\put(105.,80.){\makebox(0,0)[cc]{$ \sss - $}}
\put(114.,80.){\makebox(0,0)[cc]{$ \sss - $}}
\put(123.,80.){\makebox(0,0)[cc]{$ \sss - $}}
\put(132.,80.){\makebox(0,0)[cc]{$ \sss - $}}
\put(141.,80.){\makebox(0,0)[cc]{$ \sss - $}}
\put(150.,80.){\makebox(0,0)[cc]{$ \sss - $}}
\put(159.,80.){\makebox(0,0)[cc]{$ \sss - $}}
\put(168.,80.){\makebox(0,0)[cc]{$ \sss - $}}
\put(177.,80.){\makebox(0,0)[cc]{$ \sss - $}}
\put(186.,80.){\makebox(0,0)[cc]{$ \sss - $}}
\put(195.,80.){\makebox(0,0)[cc]{$ \sss - $}}

\put(100.,-80.){\bf \line(0,1){80.}}
\put(100.,80.){\bf \line(0,1){60.}}
\put(200.,-140.){\bf \line(0,1){60.}}
\put(200., 0.){\bf \line(0,1){80.}}

\put(100.00,-140.00){\circle{1.2}}
\put(100.00,-139.00){\circle{1.2}}
\put(100.00,-138.00){\circle{1.2}}
\put(100.00,-137.00){\circle{1.2}}
\put(100.00,-136.00){\circle{1.2}}
\put(100.00,-135.00){\circle{1.2}}
\put(100.00,-134.00){\circle{1.2}}
\put(100.00,-133.00){\circle{1.2}}
\put(100.00,-132.00){\circle{1.2}}
\put(100.00,-131.00){\circle{1.2}}
\put(100.00,-130.00){\circle{1.2}}
\put(100.00,-129.00){\circle{1.2}}
\put(100.00,-128.00){\circle{1.2}}
\put(100.00,-127.00){\circle{1.2}}
\put(100.00,-126.00){\circle{1.2}}
\put(100.00,-125.00){\circle{1.2}}
\put(100.00,-124.00){\circle{1.2}}
\put(100.00,-123.00){\circle{1.2}}
\put(100.00,-122.00){\circle{1.2}}
\put(100.00,-121.00){\circle{1.2}}
\put(100.00,-120.00){\circle{1.2}}
\put(100.00,-119.00){\circle{1.2}}
\put(100.00,-118.00){\circle{1.2}}
\put(100.00,-117.00){\circle{1.2}}
\put(100.00,-116.00){\circle{1.2}}
\put(100.00,-115.00){\circle{1.2}}
\put(100.00,-114.00){\circle{1.2}}
\put(100.00,-113.00){\circle{1.2}}
\put(100.00,-112.00){\circle{1.2}}
\put(100.00,-111.00){\circle{1.2}}
\put(100.00,-110.00){\circle{1.2}}
\put(100.00,-109.00){\circle{1.2}}
\put(100.00,-108.00){\circle{1.2}}
\put(100.00,-107.00){\circle{1.2}}
\put(100.00,-106.00){\circle{1.2}}
\put(100.00,-105.00){\circle{1.2}}
\put(100.00,-104.00){\circle{1.2}}
\put(100.00,-103.00){\circle{1.2}}
\put(100.00,-102.00){\circle{1.2}}
\put(100.00,-101.00){\circle{1.2}}
\put(100.00,-100.00){\circle{1.2}}
\put(100.00,-99.00){\circle{1.2}}
\put(100.00,-98.00){\circle{1.2}}
\put(100.00,-97.00){\circle{1.2}}
\put(100.00,-96.00){\circle{1.2}}
\put(100.00,-95.00){\circle{1.2}}
\put(100.00,-94.00){\circle{1.2}}
\put(100.00,-93.00){\circle{1.2}}
\put(100.00,-92.00){\circle{1.2}}
\put(100.00,-91.00){\circle{1.2}}
\put(100.00,-90.00){\circle{1.2}}
\put(100.00,-89.00){\circle{1.2}}
\put(100.00,-88.00){\circle{1.2}}
\put(100.00,-87.00){\circle{1.2}}
\put(100.00,-86.00){\circle{1.2}}
\put(100.00,-85.00){\circle{1.2}}
\put(100.00,-84.00){\circle{1.2}}
\put(100.00,-83.00){\circle{1.2}}
\put(100.00,-82.00){\circle{1.2}}
\put(100.00,-81.00){\circle{1.2}}
\put(100.00,-80.00){\circle{1.2}}

\put(100.00, 1.00){\circle{1.}}
\put(100.00, 2.00){\circle{1.2}}
\put(100.00, 3.00){\circle{1.2}}
\put(100.00, 4.00){\circle{1.2}}
\put(100.00, 5.00){\circle{1.2}}
\put(100.00, 6.00){\circle{1.2}}
\put(100.00, 7.00){\circle{1.2}}
\put(100.00, 8.00){\circle{1.2}}
\put(100.00, 9.00){\circle{1.2}}
\put(100.00,10.00){\circle{1.2}}
\put(100.00,11.00){\circle{1.2}}
\put(100.00,12.00){\circle{1.2}}
\put(100.00,13.00){\circle{1.2}}
\put(100.00,14.00){\circle{1.2}}
\put(100.00,15.00){\circle{1.2}}
\put(100.00,16.00){\circle{1.2}}
\put(100.00,17.00){\circle{1.2}}
\put(100.00,18.00){\circle{1.2}}
\put(100.00,19.00){\circle{1.2}}
\put(100.00,20.00){\circle{1.2}}
\put(100.00,21.00){\circle{1.2}}
\put(100.00,22.00){\circle{1.2}}
\put(100.00,23.00){\circle{1.2}}
\put(100.00,24.00){\circle{1.2}}
\put(100.00,25.00){\circle{1.2}}
\put(100.00,26.00){\circle{1.2}}
\put(100.00,27.00){\circle{1.2}}
\put(100.00,28.00){\circle{1.2}}
\put(100.00,29.00){\circle{1.2}}
\put(100.00,30.00){\circle{1.2}}
\put(100.00,31.00){\circle{1.2}}
\put(100.00,32.00){\circle{1.2}}
\put(100.00,33.00){\circle{1.2}}
\put(100.00,34.00){\circle{1.2}}
\put(100.00,35.00){\circle{1.2}}
\put(100.00,36.00){\circle{1.2}}
\put(100.00,37.00){\circle{1.2}}
\put(100.00,38.00){\circle{1.2}}
\put(100.00,39.00){\circle{1.2}}
\put(100.00,40.00){\circle{1.2}}
\put(100.00,41.00){\circle{1.2}}
\put(100.00,42.00){\circle{1.2}}
\put(100.00,43.00){\circle{1.2}}
\put(100.00,44.00){\circle{1.2}}
\put(100.00,45.00){\circle{1.2}}
\put(100.00,46.00){\circle{1.2}}
\put(100.00,47.00){\circle{1.2}}
\put(100.00,48.00){\circle{1.2}}
\put(100.00,49.00){\circle{1.2}}
\put(100.00,50.00){\circle{1.2}}
\put(100.00,51.00){\circle{1.2}}
\put(100.00,52.00){\circle{1.2}}
\put(100.00,53.00){\circle{1.2}}
\put(100.00,54.00){\circle{1.2}}
\put(100.00,55.00){\circle{1.2}}
\put(100.00,56.00){\circle{1.2}}
\put(100.00,57.00){\circle{1.2}}
\put(100.00,58.00){\circle{1.2}}
\put(100.00,59.00){\circle{1.2}}
\put(100.00,60.00){\circle{1.2}}
\put(100.00,61.00){\circle{1.2}}
\put(100.00,62.00){\circle{1.2}}
\put(100.00,63.00){\circle{1.2}}
\put(100.00,64.00){\circle{1.2}}
\put(100.00,65.00){\circle{1.2}}
\put(100.00,66.00){\circle{1.2}}
\put(100.00,67.00){\circle{1.2}}
\put(100.00,68.00){\circle{1.2}}
\put(100.00,69.00){\circle{1.2}}
\put(100.00,70.00){\circle{1.2}}
\put(100.00,71.00){\circle{1.2}}
\put(100.00,72.00){\circle{1.2}}
\put(100.00,73.00){\circle{1.2}}
\put(100.00,74.00){\circle{1.2}}
\put(100.00,75.00){\circle{1.2}}
\put(100.00,76.00){\circle{1.2}}
\put(100.00,77.00){\circle{1.2}}
\put(100.00,78.00){\circle{1.2}}
\put(100.00,79.00){\circle{1.2}}
\put(100.00,80.00){\circle{1.2}}

\put(200.00,-80.00){\circle{1.2}}
\put(200.00,-79.00){\circle{1.2}}
\put(200.00,-78.00){\circle{1.2}}
\put(200.00,-77.00){\circle{1.2}}
\put(200.00,-76.00){\circle{1.2}}
\put(200.00,-75.00){\circle{1.2}}
\put(200.00,-74.00){\circle{1.2}}
\put(200.00,-73.00){\circle{1.2}}
\put(200.00,-72.00){\circle{1.2}}
\put(200.00,-71.00){\circle{1.2}}
\put(200.00,-70.00){\circle{1.2}}
\put(200.00,-69.00){\circle{1.2}}
\put(200.00,-68.00){\circle{1.2}}
\put(200.00,-67.00){\circle{1.2}}
\put(200.00,-66.00){\circle{1.2}}
\put(200.00,-65.00){\circle{1.2}}
\put(200.00,-64.00){\circle{1.2}}
\put(200.00,-63.00){\circle{1.2}}
\put(200.00,-62.00){\circle{1.2}}
\put(200.00,-61.00){\circle{1.2}}
\put(200.00,-60.00){\circle{1.2}}
\put(200.00,-59.00){\circle{1.2}}
\put(200.00,-58.00){\circle{1.2}}
\put(200.00,-57.00){\circle{1.2}}
\put(200.00,-56.00){\circle{1.2}}
\put(200.00,-55.00){\circle{1.2}}
\put(200.00,-54.00){\circle{1.2}}
\put(200.00,-53.00){\circle{1.2}}
\put(200.00,-52.00){\circle{1.2}}
\put(200.00,-51.00){\circle{1.2}}
\put(200.00,-50.00){\circle{1.2}}
\put(200.00,-49.00){\circle{1.2}}
\put(200.00,-48.00){\circle{1.2}}
\put(200.00,-47.00){\circle{1.2}}
\put(200.00,-46.00){\circle{1.2}}
\put(200.00,-45.00){\circle{1.2}}
\put(200.00,-44.00){\circle{1.2}}
\put(200.00,-43.00){\circle{1.2}}
\put(200.00,-42.00){\circle{1.2}}
\put(200.00,-41.00){\circle{1.2}}
\put(200.00,-40.00){\circle{1.2}}
\put(200.00,-39.00){\circle{1.2}}
\put(200.00,-38.00){\circle{1.2}}
\put(200.00,-37.00){\circle{1.2}}
\put(200.00,-36.00){\circle{1.2}}
\put(200.00,-35.00){\circle{1.2}}
\put(200.00,-34.00){\circle{1.2}}
\put(200.00,-33.00){\circle{1.2}}
\put(200.00,-32.00){\circle{1.2}}
\put(200.00,-31.00){\circle{1.2}}
\put(200.00,-30.00){\circle{1.2}}
\put(200.00,-29.00){\circle{1.2}}
\put(200.00,-28.00){\circle{1.2}}
\put(200.00,-27.00){\circle{1.2}}
\put(200.00,-26.00){\circle{1.2}}
\put(200.00,-25.00){\circle{1.2}}
\put(200.00,-24.00){\circle{1.2}}
\put(200.00,-23.00){\circle{1.2}}
\put(200.00,-22.00){\circle{1.2}}
\put(200.00,-21.00){\circle{1.2}}
\put(200.00,-20.00){\circle{1.2}}
\put(200.00,-19.00){\circle{1.2}}
\put(200.00,-18.00){\circle{1.2}}
\put(200.00,-17.00){\circle{1.2}}
\put(200.00,-16.00){\circle{1.2}}
\put(200.00,-15.00){\circle{1.2}}
\put(200.00,-14.00){\circle{1.2}}
\put(200.00,-13.00){\circle{1.2}}
\put(200.00,-12.00){\circle{1.2}}
\put(200.00,-11.00){\circle{1.2}}
\put(200.00,-10.00){\circle{1.2}}
\put(200.00,-9.00){\circle{1.2}}
\put(200.00,-8.00){\circle{1.2}}
\put(200.00,-7.00){\circle{1.2}}
\put(200.00,-6.00){\circle{1.2}}
\put(200.00,-5.00){\circle{1.2}}
\put(200.00,-4.00){\circle{1.2}}
\put(200.00,-3.00){\circle{1.2}}
\put(200.00,-2.00){\circle{1.2}}
\put(200.00,-1.00){\circle{1.2}}
\put(200.00,-0.00){\circle{1.2}}

\put(200.00,140.00){\circle{1.2}}
\put(200.00,139.00){\circle{1.2}}
\put(200.00,138.00){\circle{1.2}}
\put(200.00,137.00){\circle{1.2}}
\put(200.00,136.00){\circle{1.2}}
\put(200.00,135.00){\circle{1.2}}
\put(200.00,134.00){\circle{1.2}}
\put(200.00,133.00){\circle{1.2}}
\put(200.00,132.00){\circle{1.2}}
\put(200.00,131.00){\circle{1.2}}
\put(200.00,130.00){\circle{1.2}}
\put(200.00,129.00){\circle{1.2}}
\put(200.00,128.00){\circle{1.2}}
\put(200.00,127.00){\circle{1.2}}
\put(200.00,126.00){\circle{1.2}}
\put(200.00,125.00){\circle{1.2}}
\put(200.00,124.00){\circle{1.2}}
\put(200.00,123.00){\circle{1.2}}
\put(200.00,122.00){\circle{1.2}}
\put(200.00,121.00){\circle{1.2}}
\put(200.00,120.00){\circle{1.2}}
\put(200.00,119.00){\circle{1.2}}
\put(200.00,118.00){\circle{1.2}}
\put(200.00,117.00){\circle{1.2}}
\put(200.00,116.00){\circle{1.2}}
\put(200.00,115.00){\circle{1.2}}
\put(200.00,114.00){\circle{1.2}}
\put(200.00,113.00){\circle{1.2}}
\put(200.00,112.00){\circle{1.2}}
\put(200.00,111.00){\circle{1.2}}
\put(200.00,110.00){\circle{1.2}}
\put(200.00,109.00){\circle{1.2}}
\put(200.00,108.00){\circle{1.2}}
\put(200.00,107.00){\circle{1.2}}
\put(200.00,106.00){\circle{1.2}}
\put(200.00,105.00){\circle{1.2}}
\put(200.00,104.00){\circle{1.2}}
\put(200.00,103.00){\circle{1.2}}
\put(200.00,102.00){\circle{1.2}}
\put(200.00,101.00){\circle{1.2}}
\put(200.00,100.00){\circle{1.2}}
\put(200.00,99.00){\circle{1.2}}
\put(200.00,98.00){\circle{1.2}}
\put(200.00,97.00){\circle{1.2}}
\put(200.00,96.00){\circle{1.2}}
\put(200.00,95.00){\circle{1.2}}
\put(200.00,94.00){\circle{1.2}}
\put(200.00,93.00){\circle{1.2}}
\put(200.00,92.00){\circle{1.2}}
\put(200.00,91.00){\circle{1.2}}
\put(200.00,90.00){\circle{1.2}}
\put(200.00,89.00){\circle{1.2}}
\put(200.00,88.00){\circle{1.2}}
\put(200.00,87.00){\circle{1.2}}
\put(200.00,86.00){\circle{1.2}}
\put(200.00,85.00){\circle{1.2}}
\put(200.00,84.00){\circle{1.2}}
\put(200.00,83.00){\circle{1.2}}
\put(200.00,82.00){\circle{1.2}}
\put(200.00,81.00){\circle{1.2}}
\put(200.00,80.00){\circle{1.2}}

\end{picture}

\vspace{6cm}
A more detailed analysis of the model can be made using the fact that 
there exists an exact solution. 
The energy levels of this model take the form
\be E=4JN-3JL+\sum_{j=1}^N 2\cos k_j \label{nrg}\ee 
where the variables $k_j$ are solutions of the following Bethe ansatz 
equations 
\bea -(-1)^N\exp(iLk_j)&=&\prod_{l=1}^M \frac{\sin k_j -u_l+iJ}
{\sin k_j-u_l -iJ},~~~~j=1,2,...,N \nn \\
\prod_{j=1}^N\frac{\sin k_j-u_l+iJ}{\sin k_j-u_l-iJ} &=& 
-\prod_{k=1}^M\frac{u_l-u_k-2iJ}{u_l-u_k+2iJ}, ~~~~~l=1,2,...,M. 
\label{bae} 
\eea 
The states associated with solutions of the above equations are 
eigenstates of the  total spin operator 
\be \frac 12\sum_{i=1}\l(\s^z_i+\t^z_i\r) \label{spin} \ee 
with eigenvalues $N-2M$. 

The existence of the critical point is
evident from the Bethe ansatz equations. 
Exact diagonalization of the two-site Hamiltonian shows that 
there is a unique ground state for $J>1/2$ which is given by the product 
of the two rung singlets. For $L$ sites it then follows that when $J>1/2$ 
the ground state is still the product of rung singlets with energy 
$E=-3JL$. 
This is in fact the reference state used in the Bethe 
ansatz calculation and corresponds to the case $M=N=0$ for the 
Bethe ansatz equations. 
To describe an elementary excitation to a spin 1 state  we take N=1, M=0
in (\ref{bae})  
which yields real solutions for the variable $k$, viz.
$$k=\frac{2\pi r }{L},~~~~~~~r=0,1,2,...,L-1.$$
It is then apparent from the energy 
expression (\ref{nrg}) that for $J>1/2$  these elementary excitations 
are gapped. 
We therefore deduce  that $J_c= 1/2$ gives 
the critical point 
between the gapped and gapless phases of the elementary spin excitations
alluded to earlier. 

The model also exhibits elementary bound state 
excitations which we illustrate in the two-site case. For 
$L=2,\,N=2,\,M=1$ there is a solution of the Bethe ansatz equations given by
$$u=0, ~~~~k_1=-k_2=\arccos (-J) $$ 
which describes an excited state of energy $E=-2J$. From the 
eigenvalue expression for 
(\ref{spin}) we see that this state has zero spin. Such a state has the 
interpretation of the excitation of two bound quasi-particles of 
opposite spin.

In order to obtain the solution of this model, 
we begin with the coupled spin version of the Hubbard model as introduced 
by Shastry \cite{s} with the imposition of twisted boundary conditions.
The local Hamiltonian has the form
\bea h_{i(i+1)}&=&-\s^+_i\s^-_{(i+1)}-\s^+_{(i+1)}\s^-_i-
\t^+_i\t^-_{(i+1)}-\t^+_{(i+1)}\t^-_i \nn \\
&&~~~-\frac{U}{8}
\l((\s^z_i+I)(\t^z_i+I)+(\s^z_{(i+1)}+I)(\t^z_{(i+1)}+I)\r)+\frac{U}{4}
\nn \eea
where $\{\s^\pm_i,\s^z_i\}$ and $\{\t^\pm_i,\t^z_i\}$ are
two commuting sets of Pauli matrices acting on the site $i$.
For our convenience an additional applied magnetic field term
has  been
added and an overall factor of -1 included.
For the twisted boundary term we take
\bea h_{L1}&=&-e^{-i\phi_1}\s^+_L\s^-_1-e^{i\phi_1}\s^+_L\s^-_1-
e^{-i\phi_2}\t^+_L\t^-_1-e^{i\phi_2}\t^+_L\t^-_1 \nn \\
&&~~~-\frac{U}{8}
\l((\s^z_L+I)(\t^z_L+I)+(\s^z_1+I)(\t^z_1+I)\r)+\frac{U}{4}.
\nn \eea

The first step is to apply a non-local transformation given by
\bea \th(\s^{\pm}_i)&=&\s^{\pm}_i\prod_{k=1}^{i-1}\t_k^z, \nn \\
\th(\s^z_i)&=&\s^z_i, \nn \\
\th(\t^{\pm}_i)&=&\t^{\pm}_i\prod_{k=i+1}^{L}\s_k^z, \nn \\
\th(\t^z_i)&=&\t^z_i. \nn
\eea
Under the transformation $\th$ we yield a new Hamiltonian of the form
(\ref{ham}) where the bulk two-site operators now read
\bea h_{i(i+1)}&=&-\s^+_i\s^-_{(i+1)}\t_i^z-\s^+_{(i+1)}\s^-_i\t^z_i
-\t^+_i\t^-_{(i+1)}\s^z_{(i+1)}-\t^+_{(i+1)}\t^-_i\s^z_{(i+1)} \nn \\
&&~~~-\frac{U}{8}
\l((\s^z_i+I)(\t^z_i+I)+(\s^z_{(i+1)}+I)(\t^z_{(i+1)}+I)\r)
+\frac U4 \label{ham2} \eea
and the boundary term is given by
\bea h_{L1}&=&-e^{-\phi_1}\s^+_L\s^-_1\t_L^z\prod_{k=1}^L\t^z_k
-e^{i\phi_1}\s^+_1\s^-_L\t^z_L\prod_{k=1}^L\t^z_k
-e^{-i\phi_2}\t^+_L\t^-_1\s^z_1\prod_{k=1}^L\s^z_k
-e^{i\phi_2}\t^+_1\t^-_L\s^z_1\prod_{k=1}^L\s^z_k\nn \\
&&~ -\frac{U}{8}
\l((\s^z_L+I)(\t^z_L+I)+(\s^z_1+I)(\t^z_1+I)\r)+\frac U4 \label{ham3} \eea

An important observation to make is that the boundary term above has
non-local terms. To accommodate for this, note that we may write
\bea \prod_{k=1}^L\s^z_k&=&(-1)^{\M} \nn \\
\prod_{k=1}^L\t^z_k&=&(-1)^{\N-\M} \nn \eea
where ${\M}=\sum_{i-1}^L m_i,\,{\N}=\sum_{i=1}^L n_i$ and
\bea
m&=&\frac 12(I-\s^z) \nn \\
n&=&I-\frac 12(\s^z+\t^z).  \nn \eea
Since the global operators $\M,~\N$ are conserved
quantities, we can treat the twisted boundary conditions in (\ref{ham3})
in a sector dependent manner. Letting $M$ and $N$ denote the eigenvalues
of
$\M,~\N$ respectively, we now choose
\be e^{i\phi_1}=(-1)^{(N-M)}, ~~~~e^{i\phi_2}=(-1)^M. \label{mn}\ee
The validity of making this choice without destroying the solvability
stems from the fact that states with differing values of $M$ and $N$
are orthogonal {\it independent} of the values of $\phi_1$ and $\phi_2$.
Hence we may choose different values of $\phi_1$ and $\phi_2$
for each of the subspaces corresponding to a fixed $M$ and $N$.

The next step is to now employ a local transformation on the Pauli matrices
which has the form
\bea \p(\s^{\pm})&=&\frac{1}{\sqrt 2}(\s^{\pm}+\t^{\pm}\s^z) \nn \\
\p(\s^z)&=&-\s^+\t^--\s^-\t^++\frac12(\s^z+\t^z) \nn \\
\p(\t^{\pm})&=&\frac{1}{\sqrt 2}(\t^{\pm}-\s^{\pm}\t^z) \nn \\
\p(\t^z)&=&-\s^+\t^--\s^-\t^+-\frac12(\s^z+\t^z). \nn \eea
It is worth noting that the above transformation can be expressed
$$\p(x)=TxT^{-1}$$
where
$$T=\l(\frac12+\frac{1}{2\sqrt 2}\r)\t^x-i\l(\frac12
-\frac{1}{2\sqrt 2}\r)\s^z\t^y
-\frac{i}{2\sqrt 2}\s^y-\frac{1}{2\sqrt 2}\s^x\t^z$$
is a unitary operator. Applying this transformation to the local
Hamiltonians (\ref{ham2},\ref{ham3}) gives us the local ladder Hamiltonians
$$h_{i(i+1)}=\t^+_i\s^-_{i+1}+\t^-_i\s^+_{i+1}+\l(\s^+_i\t^-_{i+1}+
\s^-_i\t^+_{i+1}\r)\t_i^z\s^z_{i+1}
+\frac{U}{8}\l({\vec \s}_i.{\vec\t}_i
+{\vec \s}_{i+1}.{\vec \t}_{i+1}\r)$$
and the global Hamiltonian has regular periodic boundary conditions.

The final step in obtaining (\ref{ham}) is to set $J=U/4$ and
perform the transformation
\bea \s_i&=&\s_i +1/2\l(1-(-1)^i\r)(\t_i-\s_i) \nn \\
\t_i&=&\t_i +1/2\l(1-(-1)^i\r)(\s_i-\t_i)
\nn \eea
 which has the effect of interchanging the leg spaces on the odd
rungs while leaving the even numbered rungs unchanged.

The energy expression for the Shastry model
with twisted boundary conditions can be obtained through
the Bethe ansatz. The result is
$$E=\frac{U(4N-3L)}{4} +\sum_{j=1}^N 2\cos k_j $$
such that the $k_j$ satisfy the Bethe ansatz equations
\bea (-1)^{M+N}e^{-i\phi_2}\exp(iLk_j)&=& -\prod^M_{l=1}
\frac{\sin k_j-u_l+iU/4}{\sin k_j -u_l-iU/4}
,~~~~j=1,2,...,N \nn \\
\prod_{j=1}^N\frac{\sin k_j -u_l+iU/4}{\sin k_j-u_l-iU/4}&=&
-(-1)^Ne^{-i(\phi_1-\phi_2)}\prod_{k=1}^M\frac{u_l-u_k-iU/2}{u_l-u_k+iU/2},~~~~~l=1,2,...,M. \nn
\eea

An important point here is that the numbers $M$ and $N$ above have 
precisely the same meaning as the interpretation presented earlier;
i.e. they are the eigenvalues of the conserved operators $\M$ and $\N$. 
Consequently, we need only substitute the values of eq. (\ref{mn}) into the 
above energy expression and Bethe ansatz equations which gives us 
(\ref {nrg},\ref{bae}) with the parametrization $J=U/4$. 

\begin{flushleft}
{\bf Acknowledgements}
\end{flushleft}
We thank Murray Batchelor, Jan de Gier, Huan-Qiang Zhou for discussions
and CNPq- Conselho Nacional de Desenvolvimento Cient\'{\i}fico e 
Tecnol\'ogico and the Australian Research Council for financial support. 
Our appreciation also to the organizers of the International 
Conference on Spin
 Ladders and Low-Dimensional Correlated Systems held at the International
Centre of Condensed Matter Physics, Brasilia, where part of this work
was completed.

\end{document}